\def\simless{\mathbin{\lower 3pt\hbox
             {$\rlap{\raise 5pt\hbox{$\char'074$}}\mathchar"7218$}}}    

\def\simmore{\mathbin{\lower 3pt\hbox
             {$\rlap{\raise 5pt\hbox{$\char'076$}}\mathchar"7218$}}}    

\documentclass[structabstract]{aa}

\usepackage{graphicx}
\usepackage{amsmath}
\usepackage{xcolor}
\begin{document}
\title{
Relativistic quantum-mechanical versus classical magnetic resonant scattering cross sections
}

\subtitle{}

\author{N. A. Loudas\inst{1,2} N. D. Kylafis\inst{1,2} and J. E. Tr\"{u}mper\inst{3}}

\institute{
University of Crete, Department of Physics \& Institute of
Theoretical \& Computational Physics, 70013 Herakleio, Greece\\
\and 
Institute of Astrophysics,
Foundation for Research and Technology-Hellas, 71110 Heraklion, Crete, Greece\\
\and
Max-Planck-Institut f\"{u}r extraterrestrische Physik, 
Postfach 1312, 85741 Garching, Germany\\
}

\date {Received ; accepted }


\abstract 
{
Radiative transfer calculations in strong (few $\times 10^{12}$ G)
magnetic fields, observed in X-ray pulsars, require accurate resonant
differential scattering cross sections.  Such cross sections exist, 
but they are quite cumbersome.
}
{
Here we compare the classical (non-relativistic) with the 
quantum-mechanical (relativistic) 
resonant differential scattering cross sections
and offer a prescription for the use of the much simpler classical
expressions with impressively accurate results.
}
{
We have expanded the quantum-mechanical 
differential cross sections and kept terms up 
to first order in $\epsilon \equiv E/m_ec^2$ 
and $B \equiv {\cal B}/{\cal B}_{cr}$, 
where $E$ is the photon energy and ${\cal B}_{cr}$ is 
the critical magnetic field, and recovered the classical differential
cross sections plus terms that are due to spin flip, which is a pure 
quantum-mechanical phenomenon.
}
{
Adding by hand the spin-flip terms to the polarization-dependent
classical differential cross sections, we find that they 
are in excellent agreement 
with the quantum mechanical ones for all energies near resonance and
all angles. We have plotted both of them and the agreement is impressive.
}
{
We give a prescription for the use of the classical 
differential cross sections that guarantees very accurate results.
}


\authorrunning{Loudas et al. 2021}
\titlerunning{Resonant scattering cross sections} 

\maketitle


\section{Introduction}

Cyclotron lines are prominent features in the spectra of X-ray pulsars.
The first cyclotron line was discovered in Hercules X-1 
(Tr\"{u}mper et al. 1977, 1978) and since then over 35 accreting magnetic
neutron stars exhibit electron 
cyclotron lines, sometimes with harmonics (Staubert et al. 2019).
They are also called cyclotron resonance scattering features (CRSFs).

Despite the many years that have passed since the first observation of a CRSF, 
it is still unclear where these features form and with what mechanism. 
Early on, it was suggested (Basko \& Sunyaev 1976) that the CRSFs are 
produced in the radiative shock in the accretion column.  However, no
calculation has been done so far for the simultaneous production of the 
power-law spectrum and the cyclotron line in a radiative shock.  Instead, 
several calculations have been performed using a slab, illuminated from 
one side (Ventura et al. 1979; Nagel 1981; Nishimura 2008; 
Araya \& Harding 1999, Araya-Gochez \& Harding 2000; Schoenherr et al. 2007). 

Another interesting idea was given by Poutanen et al. (2013), who proposed
that the CRSF is produced by reflection of the continuum emitted at the
radiative shock on the surface of the neutron star.  Again, no radiative
transfer calculation has been reported yet on this mechanism.

A possible reason for the limited detailed calculations is the 
resonant differential cross sections.  
Not their unavailability, but their complexity.  The expressions
derived by Herold (1979), Daugherty \& Harding (1986), Bussard et al. (1986),  
Harding \& Daugherty (1991), 
Sina (1996), Gonthier et al. (2014), Mushtukov et al. (2016), and 
Schwarm (2017) for the complete, Quantum Electrodynamic,
differential Compton resonant
cross sections are quite general and, because of this, 
cumbersome and rather impractical. However, this generality is not needed
in most cases.  

Most of the cyclotron lines that have been observed so far (for a review see
Staubert et al. 2019) are at 
cyclotron energies $E_c \ll m_ec^2$, implying magnetic
fields ${\cal B} \ll {\cal B}_{cr}$, where
${\cal B}_{cr}= m_e^2 c^3/ e \hbar$ is the critical magnetic field.  
For such  cases, an expansion of the full quantum mechanical cross 
sections up to
first order in $\epsilon \equiv E/m_ec^2$ and 
$B \equiv {\cal B} / {\cal B}_{cr}$ gives simpler, but
nevertheless accurate, expressions for the necessary calculations.

Known in the literature are the much simpler classical (Thomson)
cross sections 
(Canuto, Lodenquai, \& Ruderman 1971;
Blandford \& Scharlemann 1976;
Nobili, Turolla, \& Zane 2008a).  
The 
question then arises:  can a prescription be found, such that the use
of the simple classical cross sections 
in radiative transfer calculations gives very accurate results?
The present work answers this question positively.

In \S\ 2, we discuss the polarization-dependent cross sections.
First, we give the classical ones.
Then, we discuss the quantum-mechanical cross sections, that have been
derived either with the Johnson \& Lippmann (1949) wave functions or 
with the physically more meaningful 
Sokolov \& Ternov (1968) formalism, and compare them numerically
for $E \ll m_ec^2$ and ${\cal B} \ll {\cal B}_{cr}$.
We expand to first order in the small parameters
$\epsilon$ and $B$
the cross sections given by Harding \& Daugherty (1991) and Sina (1996).
Finally, we offer a prescription with which the classical cross sections
can be used for extremely accurate results.
In \S\ 3, we summarize our work.

\section{The cross sections}

\subsection{Classical}

The classical (Thomson) differential cross sections for polarized resonant
scattering are given by  Nobili, Turolla, \& Zane (2008a; see also
Canuto, Lodenquai, \& Ruderman 1971;
Blandford \& Scharlemann 1976).
After integrating over the azimuthal angle $\phi$ we have 
$$
{ {d \sigma_{11}} \over {d\cos\theta^\prime} } = 2\pi
{ {3 \pi r_0 c} \over 8} L(\omega, \omega_r)
\cos^2\theta \cos^2 \theta^\prime,
\eqno(1a)
$$
$$
{ {d \sigma_{12}} \over {d\cos\theta^\prime} } = 2\pi
{ {3 \pi r_0 c} \over 8} L(\omega, \omega_r)
\cos^2\theta,
\eqno(1b)
$$
$$
{ {d \sigma_{21}} \over {d\cos\theta^\prime} } = 2\pi
{ {3 \pi r_0 c} \over 8} L(\omega, \omega_r)
\cos^2 \theta^\prime,
\eqno(1c)
$$
$$
{ {d \sigma_{22}} \over {d\cos\theta^\prime} } = 2\pi
{ {3 \pi r_0 c} \over 8} L(\omega, \omega_r),
\eqno(1d)
$$
where the index 1 (2) stands for the ordinary (extraordinary) mode,
$r_0$ is the classical electron radius,
$$
L(\omega, \omega_r)= 
{ {\Gamma/2\pi} \over {(\omega- \omega_r)^2 +(\Gamma/2)^2} }
\eqno(2)
$$
is the normalized Lorentz profile
with $\omega = E/\hbar$ the photon frequency, 
$\omega_r = E_r/\hbar$ the resonant frequency, which
in the non-relativistic regime is the cyclotron frequency
$\omega_c = e{\cal B}/m_ec$, 
$\Gamma = 4 e^2 \omega_c^2/ (3 m_e c^3)$ accounts for the finite
transition life-time of the excited state (e.g. Daugherty \& Ventura 1978;
Ventura 1979), and
$\theta$ and $\theta^\prime$ are 
the incident and scattered angles, respectively, with respect to the 
direction of the magnetic field $\vec {\cal B}$.

The polarization averaged differential cross section is given by
$$
{ {d \sigma} \over {d \cos\theta^\prime} } =
2 \pi { {3 \pi r_0 c} \over 16} L(\omega, \omega_r)
(1+\cos^2 \theta)(1+ \cos^2 \theta^\prime).
\eqno(3)
$$

\subsection{Relativistic quantum mechanical}

Expressions for the differential Compton scattering cross sections
as functions of polarization, energy $E$, and magnetic field $\cal B$
have been derived by Harding \& Daugherty (1991),
using the Johnson \& Lippmann (1949) wavefunctions. Similar, but more
accurate, expressions have been derived by Sina (1996) using the Sokolov
\& Ternov formalism (for a detailed discussion see 
Gonthier et al. 2014). At large values of $\cal B$, there are 
differences between the two (Schwarm 2017). 

Since we are interested
at non-relativistic energies and sub-critical magnetic fields, we compare
numerically the expressions of Harding \& Daugherty (1991) with those 
of Sina (1996) for $B = 0.03 $.  
In Fig. 1, we compare $d\sigma_{11}/d\cos\theta^\prime$ near resonance for four values of the
incident angle $\theta$ with respect to the magnetic field, and
four values of the scattered angle $\theta^\prime$.  
The black lines correspond to Harding \& Daugherty (1991), while the red ones
to Sina (1996). The curves essentially overlap, but for a detailed comparison {{\it at resonance}} we show in Fig. 2 the ratio of the two curves for the cases displayed in Fig. 1.  The differences at resonance are less than 3\%.
We remark that similar agreement exists for the other
three differential cross sections.

\begin{figure}[h]
	\centering
	\includegraphics[angle=0,width=9.0cm]{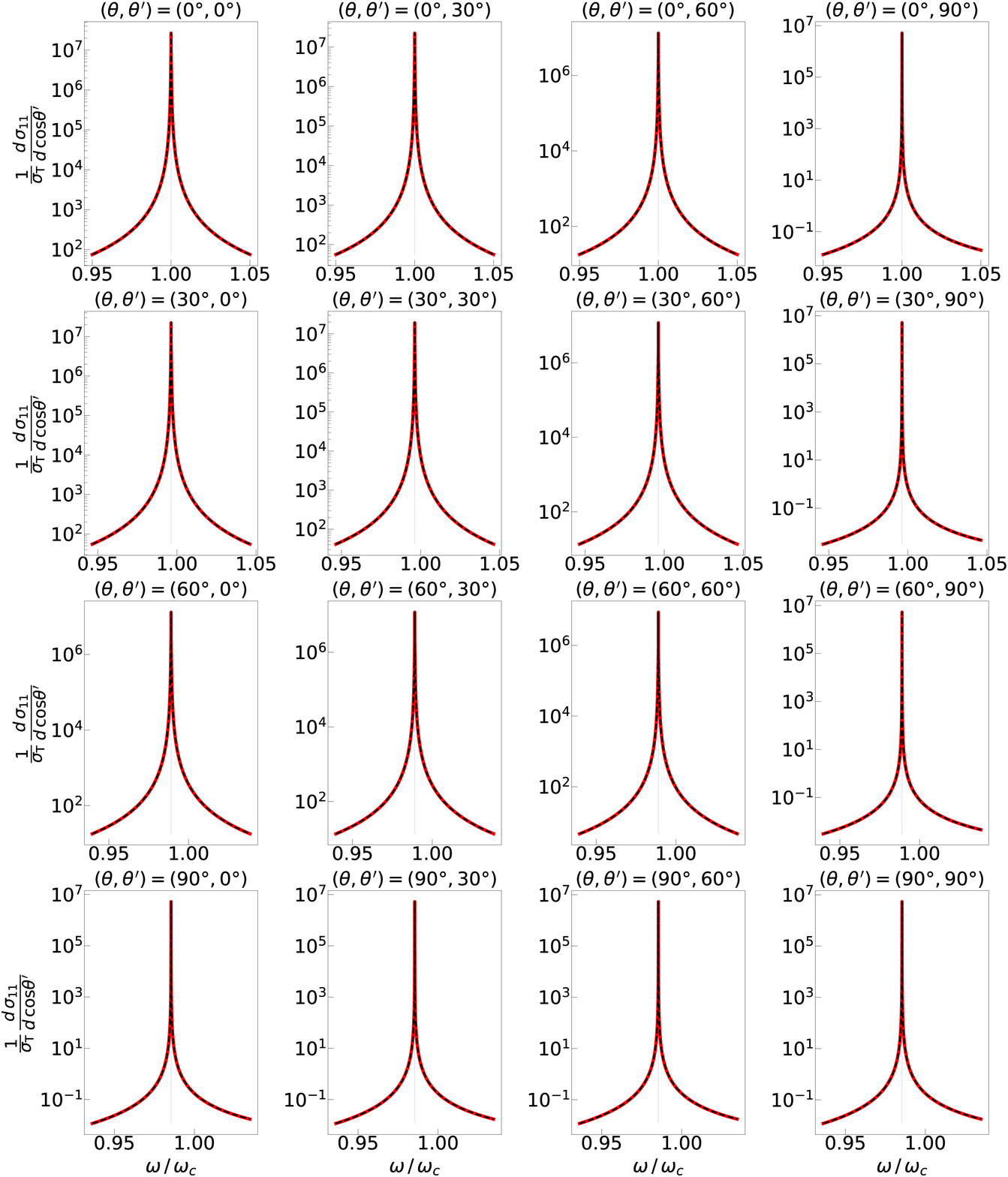}
	\caption{
		Polarization-dependent differential cross section 
		$(1/\sigma_T) ~ d\sigma_{11}/d\cos\theta^\prime$
		as a function of photon energy $E=\hbar \omega$
		for a set of incident angles $\theta$ and scattered angles $\theta^\prime$.
		The set of angles is $0, 30, 60$, and $90$ degrees, with $\theta$
		changing vertically and $\theta^\prime$ horizontally.
		The red solid lines are produced
		from the expression of Sina (1996), while the black dashed ones are produced from 
		the expression of Harding \& Daugherty (1991).
		In all cases, the resonant frequency is given by expression (7).
		The vertical line indicates the resonant frequency $\omega_r$,
		which for $\theta \ne 0$ is smaller than $\omega_c$.
		Here $E_c=\hbar \omega_c = 15.33$ keV.
	        }
	\label{Fig1}
\end{figure}

\begin{figure}[h]
	\centering
	\includegraphics[angle=0,width=9.0cm]{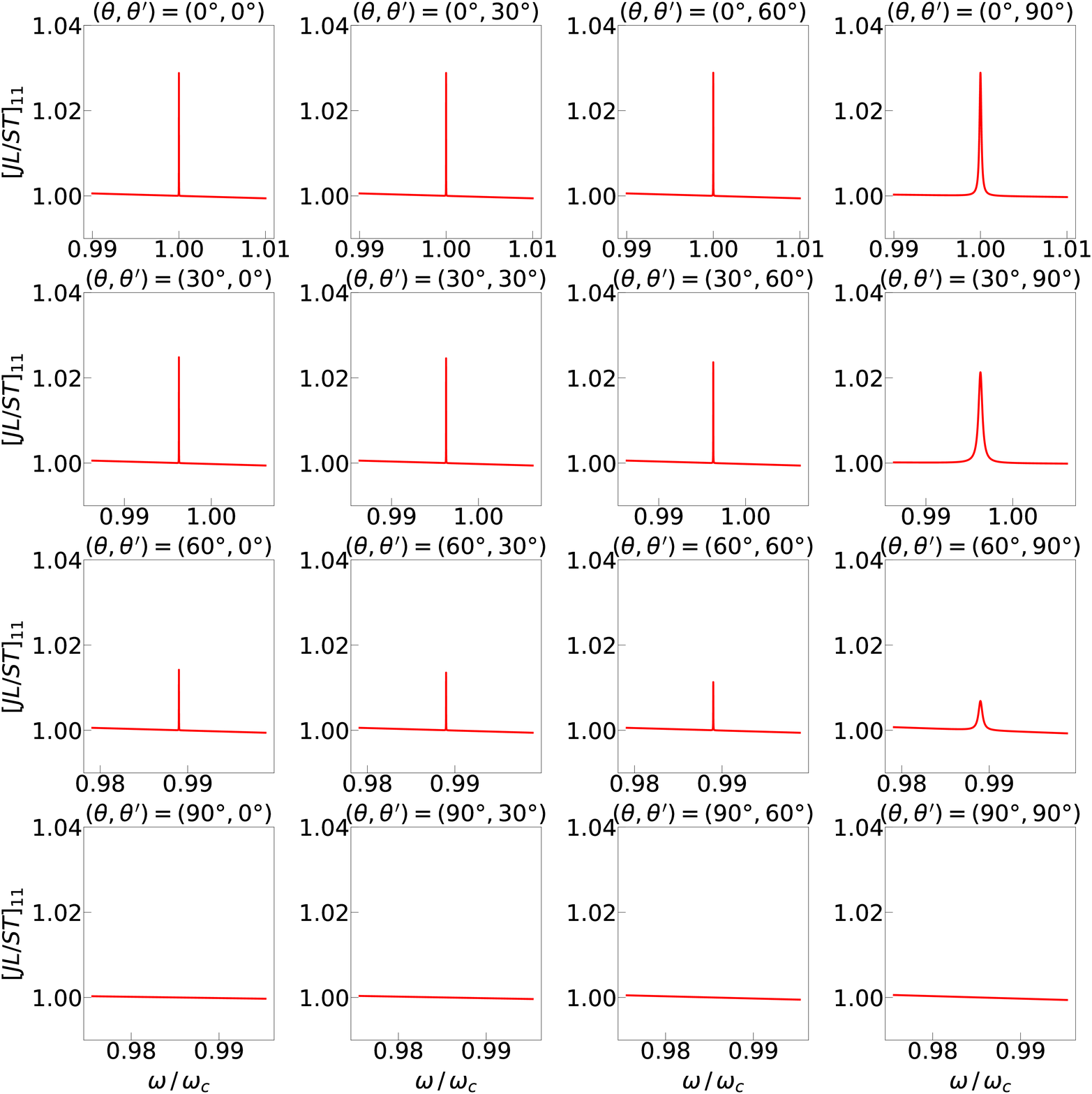}
	\caption{
		Ratio of the Harding \& Daugherty (1991) polarization-dependent differential cross section 
		$(1/\sigma_T) ~ d\sigma_{11}/d\cos\theta^\prime$ to those of Sina (1996)
		for the cases shown in Fig. 1. }
	\label{Fig2}
\end{figure}

In Fig. 3, we show in the form of heat maps the ratios {\it at resonance} of the
differential cross sections calculated from the expressions of Harding \&
Daugherty (1991) to those of Sina (1996) for $B=0.03$. 
It is evident that, for all incident angles $\theta$ and all scattered angles
$\theta^\prime$, the ratio is between 1.000 and 1.027.  Thus, for $B \ll 1$,
it is irrelevant which of the two formalisms is used.

\begin{figure}[h]
	\centering
	\includegraphics[angle=0,width=9.0cm]{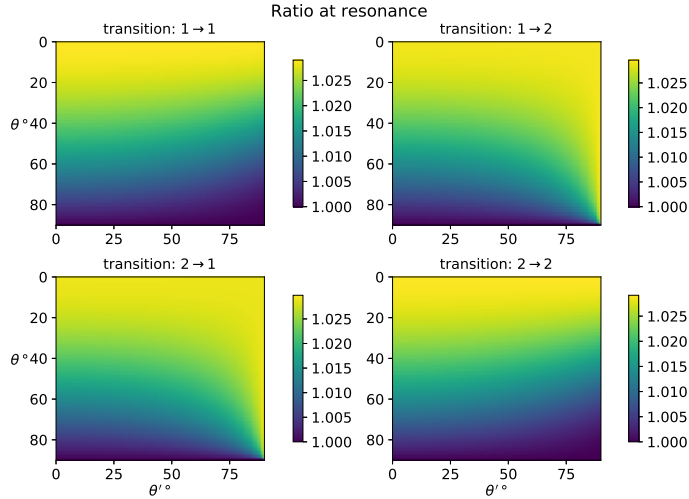}
	\caption{
		Ratio of the Harding \& Daugherty (1991) polarization-dependent
		differential cross sections at resonance to those of Sina (1996) for different 
		incident and scattered angles $\theta$ and $ \theta^\prime$, respectively. 
		In all cases, the resonant frequency is given by expression (7) and $B=0.03$. 
		}
	\label{Fig3}
\end{figure}

In order to find simpler expressions, that are nevertheless quite accurate
for $E \ll m_ec^2$ and ${\cal B} \ll {\cal B}_{cr}$, we have expanded the
expressions of Harding \& Daugherty (1991)
to first order  in the small parameters
$\epsilon=E/m_ec^2$ and $B={\cal B}/{\cal B}_{cr}$.
This is done in Appendix A.  Here we give the results.

\begin{multline*}
{ {d \sigma_{11}} \over {d\cos\theta^\prime} } = 2\pi { {3\pi r_o c} \over {8} }\Big[
g^{1 \to 1} \cdot L_{-} + h^{1 \to 1} \cdot L_{+}\\
+ \sqrt{2 B} \cos\theta\cos\theta^\prime \dfrac{L_{-} \cdot L_{+}}{L_{mix}}
\Big],
\tag{4a}
\end{multline*}
\begin{multline*}
{ {d \sigma_{12}} \over {d\cos\theta^\prime} } = 2\pi { {3\pi r_o c} \over {8} }\Big[
g^{1 \to 2} \cdot L_{-} + h^{1 \to 2} \cdot L_{+} \\
+ \sqrt{2 B} \cos\theta\cos\theta^\prime \dfrac{L_{-} \cdot L_{+}}{L_{mix}}
\Big],
\tag{4b}
\end{multline*}
\begin{multline*}
{ {d \sigma_{21}} \over {d\cos\theta^\prime} } = 2\pi { {3\pi r_o c} \over {8} }\Big[
g^{2 \to 1} \cdot L_{-} + h^{2 \to 1} \cdot L_{+} \\
+ \sqrt{2 B} \cos\theta\cos\theta^\prime \dfrac{L_{-} \cdot L_{+}}{L_{mix}}
\Big],
\tag{4c}
\end{multline*}
\begin{multline*}
{ {d \sigma_{22}} \over {d\cos\theta^\prime} } = 2\pi { {3\pi r_o c} \over {8} }\Big[
g^{2 \to 2} \cdot L_{-} + h^{2 \to 2} \cdot L_{+} \\
+ \sqrt{2 B} \cos\theta\cos\theta^\prime \dfrac{L_{-} \cdot L_{+}}{L_{mix}}
\Big],
\tag{4d}
\end{multline*}
where 
$L_{-}(\omega,\omega_r),\, L_{+}(\omega,\omega_r)$ and 
$L_{mix}(\omega,\omega_r)$ are the normalized Lorentz profiles, which are 
characterized by the decay widths $\Gamma^\prime_{-},\,\Gamma^\prime_{+}$,
and  $\Gamma^\prime_{mix}$, respectively, and are given in Appendix A. 
The resonant frequency $\omega_r$ is given in eq. (7) below.
In addition, $g^{s \to s^\prime}(\theta, \theta^\prime, B)$ and 
$h^{s \to s^\prime}(\theta, \theta^\prime, B)$ are first degree polynomials 
in $B$, and are given in Appendix A. 
Note that, $s$ represents the polarization mode of the incident photon,
whereas $s^\prime$ that of the scattered photon.
 
In Appendix B, we have expanded the expressions of Sina (1996)
to first order in $\omega=E/(\hbar m_ec^2)$ and $B={\cal B}/{\cal B}_{cr}$.
The results are:
\begin{multline*}
{ {d \sigma_{11}} \over {d\cos\theta^\prime} } = 2\pi { {3\pi r_o c} \over {8} }\Big[
{\cal{G}}^{1 \to 1} \cdot L_{-} + h^{1 \to 1} \cdot L_{+} \\
+ \sqrt{2 B} \cos\theta\cos\theta^\prime \dfrac{L_{-} \cdot L_{+}}{L_{mix}}
\Big],
\tag{5a}
\end{multline*}
\begin{multline*}
{ {d \sigma_{12}} \over {d\cos\theta^\prime} } = 2\pi { {3\pi r_o c} \over {8} }\Big[
{\cal{G}}^{1 \to 2} \cdot L_{-} + h^{1 \to 2} \cdot L_{+}\\
+ \sqrt{2 B} \cos\theta\cos\theta^\prime \dfrac{L_{-} \cdot L_{+}}{L_{mix}}
\Big],
\tag{5b}
\end{multline*}
\begin{multline*}
{ {d \sigma_{21}} \over {d\cos\theta^\prime} } = 2\pi { {3\pi r_o c} \over {8} }\Big[
{\cal{G}}^{2 \to 1} \cdot L_{-} + h^{2 \to 1} \cdot L_{+} \\
+ \sqrt{2 B} \cos\theta\cos\theta^\prime \dfrac{L_{-} \cdot L_{+}}{L_{mix}}
\Big],
\tag{5c}
\end{multline*}
\begin{multline*}
{ {d \sigma_{22}} \over {d\cos\theta^\prime} } = 2\pi { {3\pi r_o c} \over {8} }\Big[
{\cal{G}}^{2 \to 2} \cdot L_{-} + h^{2 \to 2} \cdot L_{+}\\
+ \sqrt{2 B} \cos\theta\cos\theta^\prime \dfrac{L_{-} \cdot L_{+}}{L_{mix}}
\Big],
\tag{5d}
\end{multline*}
where $L_{-}(\omega,\omega_r),\, L_{+}(\omega,\omega_r), \,
L_{mix}(\omega,\omega_r)$ and $h^{s\to s^\prime}$ are the same as in (4a)-(4d),
while the correction functions ${\cal G}^{s \to s^\prime}$ are slightly 
different from the $g^{s\to s^\prime}$ ones and are given in Appendix B. 
Note that, in order to avoid confusion between the different notations, 
we have employed the notation of Harding \& Daugherty (1991) in eqs. (5) 
instead of the notation of Sina (1996) that we systematically use in Appendix B.

It is important to notice two things: a) The similarity of eqs. (5) with 
those of (4).  They are
identical, except for the small differences between $g$ and $\cal G$. This,
of course, is not surprising given the ratio plots in Fig. 2 and the heat plots in Fig. 3.
b) The expressions in eqs. (4) and (5) are much simpler than the full
expressions given by Harding \& Daugherty (1991) and Sina (1996),
respectively.

\subsection{Even simpler expressions}

Looking at expressions (4) and (5), we have wondered whether all the terms
in them are crucial.  Thus, we have written the extremely simple
expressions that follow.

$$
{ {d \sigma_{11}} \over {d\cos\theta^\prime} } \approx 2\pi
{ {3 \pi r_0 c} \over 8} \left(
\cos^2\theta \cos^2 \theta^\prime \cdot L_{-}
+ {B \over 2} \cdot L_{+}
\right)
,
\eqno(6a)
$$
$$
{ {d \sigma_{12}} \over {d\cos\theta^\prime} } \approx 2\pi
{ {3 \pi r_0 c} \over 8} \left(
\cos^2\theta \cdot L_{-}
+ {B \over 2} \cos^2 \theta^\prime\cdot L_{+}
\right)
,
\eqno(6b)
$$
$$
{ {d \sigma_{21}} \over {d\cos\theta^\prime} } \approx 2\pi
{ {3 \pi r_0 c} \over 8} \left(
\cos^2\theta^\prime \cdot L_{-}
+ {B \over 2} \cos^2 \theta \cdot L_{+}
\right)
,
\eqno(6c)
$$
$$
{ {d \sigma_{22}} \over {d\cos\theta^\prime} } \approx 2\pi
{ {3 \pi r_0 c} \over 8} \left( L_{-}
+ {B \over 2} \cos^2\theta\cos^2 \theta^\prime\cdot L_{+}
\right)
.
\eqno(6d)
$$
The terms proportional to $L_{-}$ are identical to the classical cross sections
(1), because $L_{-}$ is equal to $L$ given by eq. (2). The terms proportional
to $L_{+}$ are non-classical, because they contain the decay width 
$\Gamma_{+}$, which is associated with the spin flip.

In Fig. 4, we compare expression (6b) with expression (5b)
for $B = 0.03$ and in Fig. 5, we present the ratio of these two formulae. The differences are impressively negligible.
The same is true for the heat maps {\it at resonance} (Fig. 6).

\begin{figure}[h]
	\centering
	\includegraphics[angle=0,width=9.0cm]{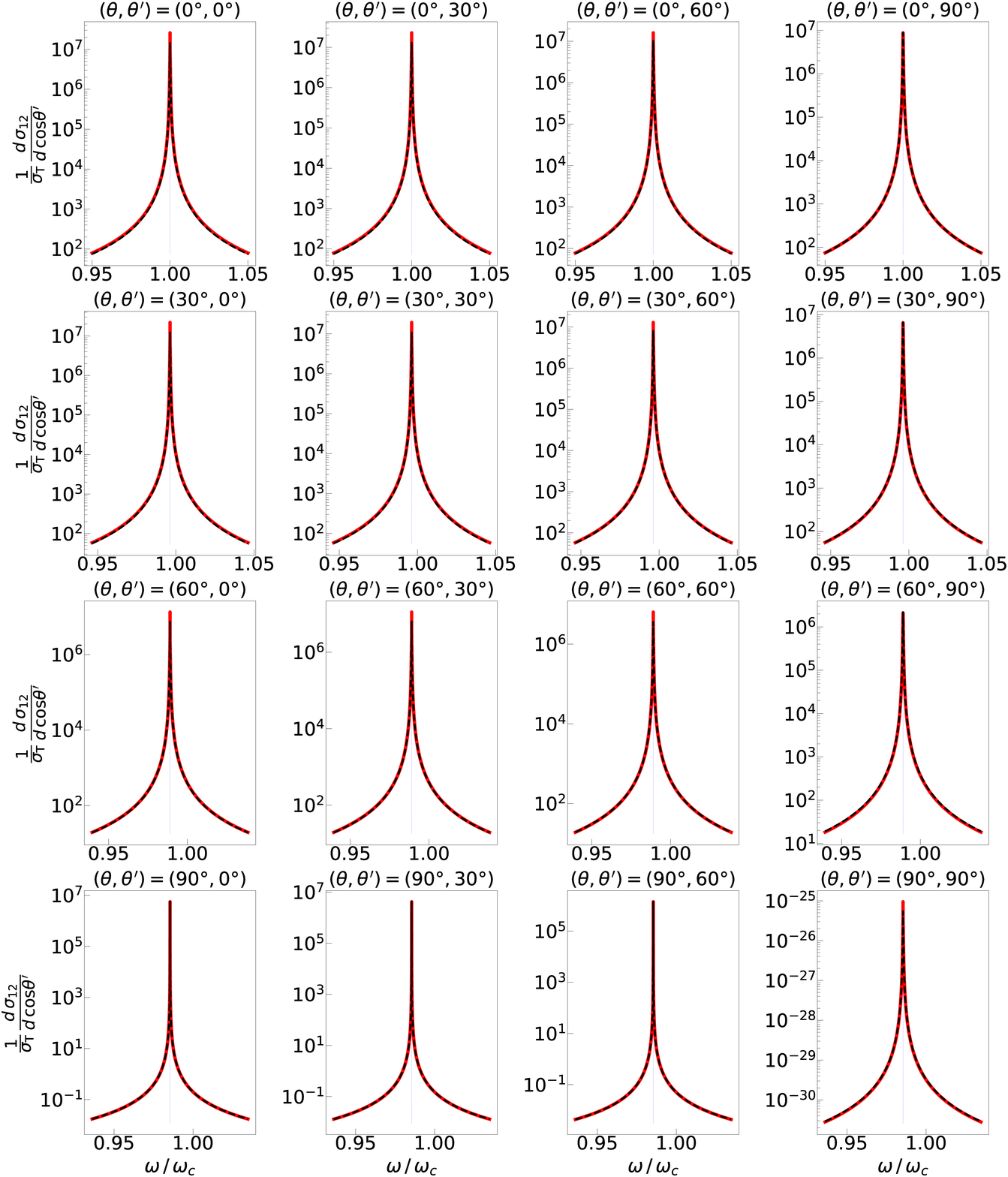}
	\caption{
		Polarization-dependent resonant differential cross section 
		$(1/\sigma_T) ~ d\sigma_{12}/d\cos\theta^\prime$
		as a function of photon energy $E=\hbar \omega$
		for a set of incident angles $\theta$ and scattered angles 
		$\theta^\prime$.
		The set of angles is $0, 30, 60$, and $90$ degrees, 
		with $\theta$
		changing vertically and $\theta^\prime$ horizontally.
		The red solid lines are produced
		from the first order expansion (5b) of Sina (1996), 
		while the black dashed ones are produced from 
		the simplified expression (6b).
		In all cases, the resonant frequency is given by expression (7).
		The vertical line indicates the resonant frequency $\omega_r$,
		which for $\theta \ne 0$ is smaller than $\omega_c$.
		Here $E_c=\hbar \omega_c = 15.33$ keV. 
		}
	\label{Fig4}
\end{figure}

\begin{figure}[h]
	\centering
	\includegraphics[angle=0,width=9.0cm]{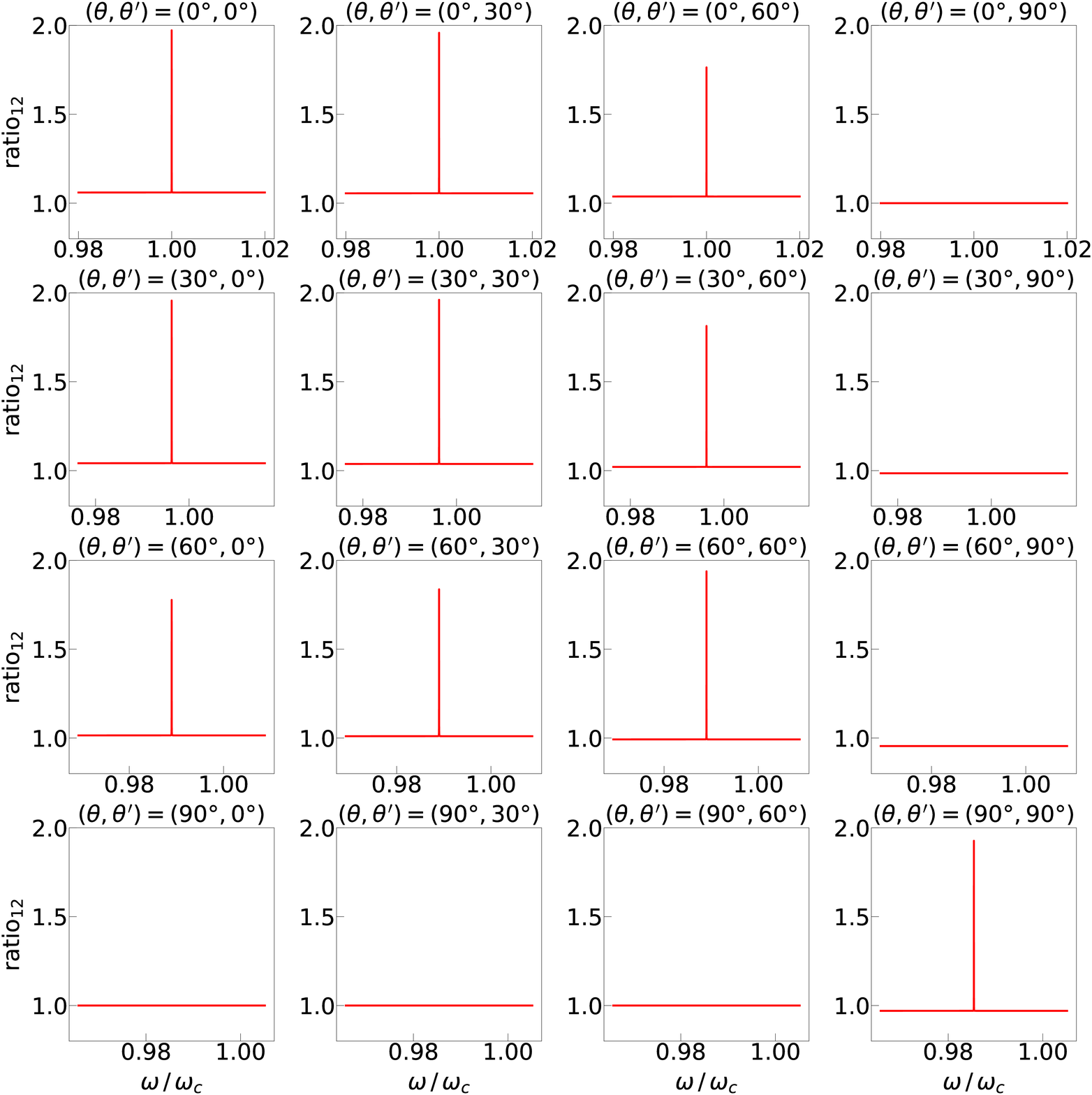}
	\caption{
		Ratio of the polarization-dependent differential cross sections $d\sigma_{12}/d\cos\theta^\prime$ given by (5b) to the simplified expression (6b) for the cases shown in Fig. 4. 
	             }
	\label{Fig5}
\end{figure}

\begin{figure}[h]
	\centering
	\includegraphics[angle=0,width=9.0cm]{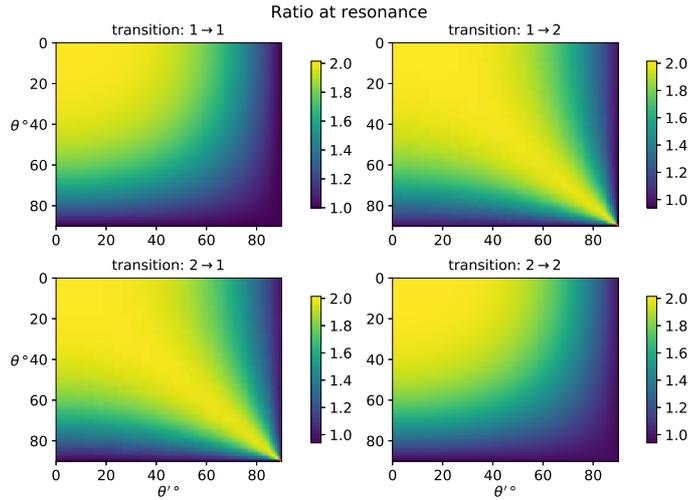}
	\caption{
		Ratio at resonance of the exact polarization-dependent 
		differential cross sections (5) to the approximate expressions (6) 
		for different incident and scattered angles $\theta$ and  $\theta^\prime$,
		respectively. In all cases, the resonant frequency is given by expression (7)
		and $B=0.03$.}
	\label{Fig6}
\end{figure}

For quick calculations with $B \ll 1$, where high accuracy is not demanded, one may safely use the very simple expressions (6), while for more accurate ones, expressions (5) are recommended.  Of course, the highest accuracy is provided by the expressions of Sina (1996).
It is our hope that these very simple expressions, together with
the prescription that we give in the next subsection, will make
resonant Compton scattering calculations easily doable.

\subsection{The prescription}

If, in a calculation, one wants to use the much simpler differential 
cross sections (1a) - (1d), then for accurate results
the following prescription should be followed.

1) The spin-flip terms must be added by hand.  Thus, eqs. (6a) - (6d)
should be used.  Of course, the exact results (5a) - (5d) are not that
complicated.

2) For the resonant frequency $\omega_r$,
the relativistically correct expression
$$
\omega_r= { { m_ec^2} \over {\hbar} }
{ { 2 B } \over
{ 1+\sqrt{1+2 B \sin^2\theta } }
}
\eqno(7)
$$ 
should be used in the Lorentz profile (2) and not the cyclotron
frequency $\omega_c= E_c/\hbar$.

3) In the classical (Thomson) limit, there is no change in the
photon energy after scattering, i.e. $\epsilon^\prime = \epsilon$. 
However, naturally there is an energy
change (see \S \ Appendix A) and the prescription dictates that 
the energy $\epsilon^\prime$ of the photon after scattering 
should be taken equal
to 
$$
\epsilon^\prime=
{ {\epsilon^2 \sin^2\theta + 2\epsilon} \over 
	{1 + \epsilon(1-\cos\theta\cos\theta^\prime)+
		\sqrt{ f_1(\epsilon, \theta, \theta^\prime) 
			+      f_2(\epsilon, \theta, \theta^\prime) } } },
\eqno(8a)
$$
where
$$
f_1(\epsilon, \theta, \theta^\prime) =
1 + 2\epsilon \cos\theta' (\cos\theta^\prime -\cos\theta),
\eqno(8b)
$$
and
$$ 
f_2(\epsilon, \theta, \theta^\prime) =
\epsilon^2(\cos\theta - \cos\theta^\prime)^2.
\eqno(8c)
$$

If the polarization of the CRSF is not of interest, then the 
polarization-averaged cross section can be used and it is 
$$
{ {d \sigma} \over {d \cos\theta^\prime} } =
2 \pi { {3 \pi r_0 c} \over 16}
(1+\cos^2 \theta)(1+ \cos^2 \theta^\prime)\left[L_{-} 
+ {B \over 2} \cdot L_{+}\right].
\eqno(9)
$$
Note that for $B \ll 1$, the polarization-averaged cross
section (3) is very accurate.  No additional terms due to spin flip are
necessary.

In this paper, we have restricted ourselves to cyclotron energies
$E_c \simless 50$ keV, because most of the observed CRSFs are in this 
energy range (Staubert et al. 2019).  However, a few CRSFs with $E_c > 50$ keV
have been observed (Staubert et al. 2019), plus the recently reported 
(Ge et al. 2020) champion with $E_c = 90.32$ keV. For this reason, 
in Fig. 7 we compare the relativistically correct
$d\sigma_{12}/d\cos\theta^\prime (\theta=0,
\theta^\prime=\pi/2)$ of Sina (1996)
with the modified classical one (eq. 6b)
for $E_c = 25$ keV (left panel), 50 keV (middle panel), and
100 keV (right panel).  Clearly, 100 keV is not much less than $m_ec^2$.
Nevertheless, we show it for the readers to see the 
magnitude of the discrepancy 
between classical and quantum-mechanical cross sections.

\begin{figure}[h]
\centering
\includegraphics[angle=0,width=9.0cm]{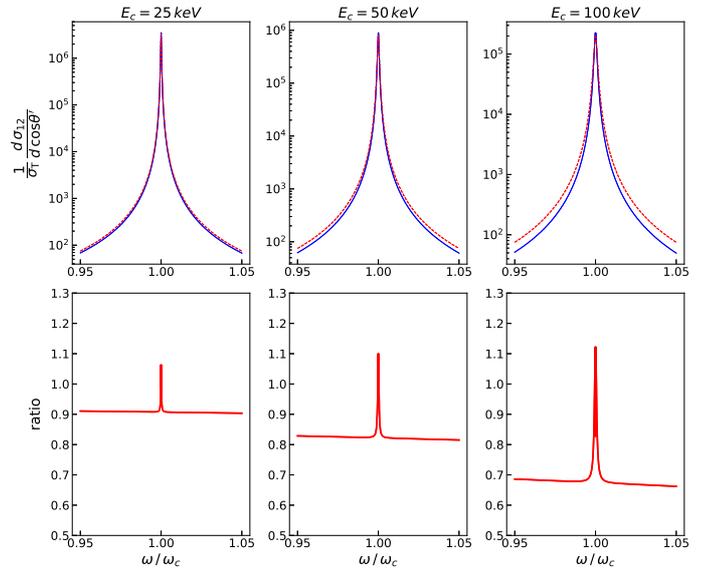}
\caption{
{\it Top panels:} Comparison of the relativistically correct
$d\sigma_{12}/d\cos\theta^\prime (\theta=0,
\theta^\prime=\pi/2)$ from Sina (1996) (blue solid line)
with the modified classical one (eq. 6b) (red dashed line)
for $E_c = 25$ keV (left panel), 50 keV (middle panel), and
100 keV (right panel).
{{\it Bottom panels:}} Corresponding ratios of the curves shown in the top panels
}
\label{Fig7}
\end{figure}

\section{Summary}

We have shown numerically that, for photon energies $E \ll mc^2$ and magnetic 
field strengths ${\cal B} \ll {\cal B}_{cr}$, the relativistic 
polarization-dependent resonant differential cross sections of Harding \& 
Daugherty (1991), derived with the Johnson \& Lippmann (1949) wave functions,
are in excellent agreement with the ones of Sina (1996), derived with the
Sokolov \& Ternov (1968) formalism.

We have shown analytically that the expansion up to first order in
$\epsilon=E/m_ec^2$ and $B={\cal B}/{\cal B}_{cr}$ of the expressions of
Harding \& Daugherty (1991) and of Sina (1996) lead to nearly identical
results.

We have provided simple, heuristic, but very accurate, expressions for
the polarization-dependent resonant differential cross sections, together
with a prescription that should be followed, for easy to perform
detailed calculations involving resonant scattering at $E_c \simless 50$ keV.

\begin{acknowledgements}

We thank an anonymous referee for pointing out to us the Sokolov \& Ternov
(1968) formalism.  
We also thank Roberto Turolla 
and Alexander Mushtukov for useful
discussions concerning the Sokolov \& Ternov (1968) formalism.
We are indebted to Alice Harding for sending us the
unpublished PhD Thesis of Ramin Sina and to Peter Gonthier for e-mail exchanges.
N.D.K. would like to thank Sterl Phinney for his notes on the 
classical magnetic scattering cross sections.
\end{acknowledgements}

\newpage 

\begin{appendix}
	
\section{Approximations resulting from the Harding \& Daugherty (1991) 
expressions}

We start from eq. (11) of Harding \& Daugherty (1991) that 
has been derived in the electron's rest frame, 
considering that the electron initially occupies 
the ground state. In a unit system where $\hbar=c=m_e=1$, 
this equation can be written as follows

\begin{multline*}
\dfrac{d\sigma_{ss^\prime}^{0,l}}{d\Omega^\prime}=\dfrac{3\sigma_T}{4}
\dfrac{\omega^\prime}{\omega}\dfrac{\exp\left(-(\omega^2\sin^2\theta + 
\omega^{\prime2}\sin^2\theta^\prime)/2B \right)}{1 + 
\omega - 
\omega^\prime - (\omega\cos\theta - 
\omega^\prime\cos\theta^\prime)\cos\theta^\prime}\\
\times\left|\sum_{n=0}^{\infty}\sum_{i=1}^{2} 
\left(F^{(1)}_{n,i}e^{i\Phi} + 
F^{(2)}_{n,i}e^{-i\Phi} \right)\right|^2,
\tag{A1}
\end{multline*}
where $\sigma_T$ is the Thomson cross section, 
$\omega$, $\omega^\prime$ are the incident 
and scattered photon frequencies in units 
of $m_ec^2/\hbar$, $\theta$ and $\theta^\prime$ 
are the incident and scattered photon angles 
with respect to the magnetic field direction, 
$B$ is the magnetic field strength $\cal B$ (in units 
of ${\cal B}_{cr}$), $\Phi$ is proportional to 
$\sin(\phi-\phi^\prime)$, $\phi$ and $\phi^\prime$ 
are the azimuthal angles of the incident and the 
scattered photon, respectively. The terms 
$F^{(1)}_{n,i}$, $F^{(2)}_{n,i}$ are complex 
functions of $\theta$, $\theta^\prime$, 
$\omega$, $\omega^\prime$, $B$, $\phi$, 
$\phi^\prime$, $s$, $s^\prime$ and can be found 
in the Appendix of Harding \& Daugherty (1991), 
where the upper indices 1, 2 are referred to 
the 1st and 2nd Feynman diagram. Furthermore, 
$s$ and $s^\prime$ stand for the incident and 
scattered photon's polarization mode and $l$ 
is the final electron's Landau state. Note that, 
the infinite sum in eq. (A1) is carried out over 
intermediate Landau states with principal quantum 
number $n$, whereas the sum over $i={1,2}$ is 
for the possible electron spin orientations 
(spin-up, spin-down) in the intermediate state.

Equation (A1) is quite general, 
but this generality is not needed in 
most cases. Specifically, for resonant 
scattering of photons with energy $E \simless m_e c^2$ 
in a magnetic field smaller than or 
comparable to the critical one ${\cal B}_{cr}= m_e^2 c^3/e\hbar$, 
Nobili, Turolla, \& Zane (2008b, hereafter 
NTZ08b) pointed out that,
for such magnetic fields, the probability 
of exciting Landau levels above
the second ($n=2$) is very small, thus the 
infinite sum over $n$ in eq. (A1) becomes finite ($n \le 2$).

Moreover, for photon energies $E \ll m_ec^2$, 
and magnetic fields ${\cal B} \ll {\cal B}_{cr}$, 
that we are interested in,  
the cross section expressions can be simplified
even more, keeping only the term $n=1$ in the sum 
over all the possible intermediate states and 
setting $l=0$ for the final electron Landau state. 
Besides that, as NTZ08b stated, the $F^{(1)}_{n=1,i}$ 
terms exhibit a divergent behavior near resonant 
frequency (see their eqs. 7 and 8), 
while the terms $F^{(2)}_{n=1,i}$, 
remain instead finite. Given that, and as long as 
we are interested in resonant scattering 
(i.e. photon frequency near the resonant one), 
the above arguments lead us to a major simplification.

Neglecting all the non-resonant terms and the 
contributions of ($n\neq1$) in eq. (A1),
we deduce the following expression
\begin{multline*}
\dfrac{d\sigma_{ss^\prime}^{0,l=0}}{d\Omega^\prime}=
\dfrac{3\sigma_T}{4}\dfrac{\omega^\prime}{\omega}
\dfrac{\exp\left(-(\omega^2\sin^2\theta + 
\omega^{\prime2}\sin^2\theta^\prime)/2B \right)}{1 + 
\omega - \omega^\prime - (\omega\cos\theta - 
\omega^\prime\cos\theta^\prime)\cos\theta^\prime}\\
\times\left|F^{(1)}_{1,1} + F^{(1)}_{1,2}\right|^2.
\tag{A2}
\end{multline*}

NTZ08b have managed to write 
the above equation in a simpler way, 
obtaining compact expressions for the 
$F^{(1)}_{1,i}$ terms. Specifically, using 
eq. (14) of NTZ08b in eq. (10) of NTZ08b one gets
\begin{multline*}
{{d\sigma_{ss^\prime}} \over {d\Omega^\prime}}=
{ {3\sigma_T} \over {16\pi} }
{ {\epsilon^\prime} \over {\epsilon} }
{{(2+\epsilon-\epsilon^\prime) 
\exp \left(
- {{\epsilon^2 \sin^2\theta + {\epsilon^\prime}^2 \sin^2\theta^\prime}
\over {2B}} \right)}
\over
{1+\epsilon-\epsilon^\prime - 
(\epsilon \cos\theta -\epsilon^\prime\cos\theta^\prime) 
\cos\theta^\prime}} \\
\times
\left({ {1+E_1}\over {2E_1} }\right)^2
\left|{ {T_+^{s \to s^\prime}}\over {1 + 
\epsilon - E_1 + i \Gamma_{+}/2} } + 
{ {T_-^{s \to s^\prime}}\over {1 + \epsilon - 
E_1 + i \Gamma_{-}/2}}  \right|^2,
\tag{A3}
\end{multline*}
where, for convenience, we retain the 
notation of
NTZ08b, but for simplicity we drop 
the indices referring to $n = 1$ and $l = 0$. 
So, $\epsilon$, $\epsilon^\prime$ are 
the incident and scattered photon 
energies in units of $m_ec^2$, $\theta$ and
$\theta^\prime$ are the incident and 
scattered photon angles with respect to 
the magnetic field direction, $B$ is 
the magnetic field strength $\cal B$
(in units of ${\cal B}_{cr}$) and $\Gamma_{+}$, $\Gamma_{-}$
are the relativistic decay rates 
corresponding to the n=1 intermediate 
state and are given by Herold, Ruder, \& Wunner (1982, 
see also Harding \& Daugherty 1991; Pavlov et al. 1991; Baring et al. 2005; NTZ08b), 
where the index "+" stands for the electron 
in the intermediate state with spin-up whereas 
the index "-" stands for spin-down. In addition, $s$ and $s^\prime$ 
refer to the incident and scattered 
photon polarization modes, and the explicit 
expressions for the $T_+^{s \to s^\prime}$, 
$T_-^{s \to s^\prime}$ terms are given 
in the Appendix of NTZ08b for $n=1$ and $l=0$.

We remark that we do not use the 
relativistic polarization-dependent resonant Compton differential 
cross section expressions of NTZ08b, because, for their purposes, they 
substituted all the Lorentz profiles with 
a $\delta-$function, and as a result the 
information about the exact shape of the 
different line profiles is lost.

The energy $E_1$ is the electron intermediate 
state and, in units of $m_ec^2$, it is given by eq. (12) of NTZ08b
(see also eq. 4 of Harding \& Daugherty 1991).
$$
E_1=\sqrt{1 + \epsilon^2\cos^2\theta + 2B}. 
\eqno(A4)
$$

The energy $\epsilon^\prime$ of 
the photon after scattering, in units 
of $m_ec^2$, is given by 
eq. (6) of NTZ08b (see also eq. 12 
of Harding \& Daugherty 1991)
$$
\epsilon'=
{ {\epsilon^2 \sin^2\theta + 2\epsilon} \over 
	{1 + \epsilon(1-\cos\theta\cos\theta^\prime)+
		\sqrt{ f_1(\epsilon, \theta, \theta^\prime) 
			+      f_2(\epsilon, \theta, \theta^\prime) } } },
\eqno(A5a)
$$
where
$$
f_1(\epsilon, \theta, \theta^\prime) =
1 + 2\epsilon \cos\theta' (\cos\theta^\prime -\cos\theta),
\eqno(A5b)
$$
and
$$ 
f_2(\epsilon, \theta, \theta^\prime) =
\epsilon^2(\cos\theta - \cos\theta^\prime)^2.
\eqno(A5c)
$$

After a lengthy but straightforward 
calculation, and a trivial integration 
over the angle $\phi^\prime$, one can write 
eq. (A3) in the following way
\begin{multline*}
{{d\sigma_{ss^\prime}} \over {d\cos\theta^\prime}}=
{ {3\pi\sigma_T} \over {16} }
{ {(1+E_1)^2} \over {E_1\sqrt{1 + 2B\sin^2\theta}}}
{ {\epsilon^\prime} \over {\epsilon} }
A \\
\times
\left[{ {(T_+^{s \to s^\prime})^2}\over {\Gamma_{+} }} {\cal L}_{+} + 
{ {(T_-^{s \to s^\prime})^2}\over {\Gamma_{-} }} {\cal L}_{-} + 
2{{T_+^{s \to s^\prime}T_-^{s \to s^\prime}} \over {\Gamma_{mix}} } 
{{{\cal L}_{+}{\cal L}_{-}} \over {{\cal L}_{mix}}} \right],
\tag{A6}
\end{multline*}
where the effective decay rates $\Gamma^e_{\pm}$, which result from the change in the Lorentz profiles argument (see NTZ08b), must be used in the Lorentz profiles (see eqs. A8-A10)
              
$$
\Gamma^e_{\pm}=\dfrac{E_1(\epsilon_r)}{\sqrt{1 + 2B\sin^2\theta}}\Gamma_{\pm}
\eqno(A7a)
$$
and we have defined the following function
$$
A=
{{(2+\epsilon-\epsilon^\prime) 
		\exp \left(
		- {{\epsilon^2 \sin^2\theta + {\epsilon^\prime}^2 \sin^2\theta^\prime}
			\over {2B}} \right)}
	\over
	{1+\epsilon-\epsilon^\prime - 
		(\epsilon \cos\theta -\epsilon^\prime\cos\theta^\prime) \cos\theta^\prime}}.
\eqno(A7b)	
$$

The quantities ${\cal L}_+(\epsilon,\epsilon_r)$, 
${\cal L}_-(\epsilon,\epsilon_r)$ \& ${\cal L}_{mix}(\epsilon,\epsilon_r)$ 
are the 
dimensionless Lorentz profiles and 
are given by
$$
{\cal L}_{+}(\epsilon, \epsilon_r)= 
{ {\Gamma^e_{+}/2\pi} \over {(\epsilon - 
\epsilon_r)^2 + (\Gamma^e_{+}/2)^2} },
\eqno(A8)
$$
$$
{\cal L}_{-}(\epsilon, \epsilon_r)= 
{ {\Gamma^e_{-}/2\pi} \over {(\epsilon - 
\epsilon_r)^2 + (\Gamma^e_{-}/2)^2} },
\eqno(A9)
$$
$$
{\cal L}_{mix}(\epsilon, \epsilon_r)= 
{ {\Gamma^e_{mix}/2\pi} \over {(\epsilon - 
\epsilon_r)^2 + (\Gamma^e_{mix}/2)^2} },
\eqno(A10)
$$
where $\Gamma^e_{mix}$ is calculated by
$$
\Gamma^e_{mix}=\dfrac{E_1(\epsilon_r)}{\sqrt{1 + 2B\sin^2\theta}}\Gamma_{mix},
\eqno(A11a)
$$
with
$$
\Gamma_{mix}=\sqrt{\Gamma_{+}\Gamma_{-}},
\eqno(A11b)
$$
and $\epsilon_r$ is the resonant energy in 
units of $m_ec^2$. It is given by eq. (8) of 
NTZ08b (see also eq. 5 of Harding \& Daugherty 1991)
$$
\epsilon_r = \epsilon_1 = 
{ {2B} \over {1+\sqrt{1+2B\sin^2\theta}} }.
\eqno(A12)
$$

Having obtained the fully relativistic 
cross sections we proceed to the 
derivation of eqs. (4a)-(4d).

Expansion up to first order 
in the small parameters $\epsilon$, $\epsilon^\prime$, and
$B$ yields
$$
{ \epsilon^\prime \over \epsilon} \approx 1-
\dfrac{\epsilon}{2}(\cos\theta - \cos\theta^\prime)^2,
\eqno(A13)
$$
$$
{
{(1+E_1)^2} \over {E_1 \sqrt{1+2B \sin^2\theta}}
}{ {\epsilon^\prime} \over {\epsilon} } A
\approx 8\left[1-B\left(2\sin^2\theta + 
(\cos\theta - \cos\theta^\prime)^2 \right) \right].
\eqno(A14)
$$
To lowest order, $\Gamma_{+}$, $\Gamma_{-}$ 
are given in Herold, Ruder, \& Wunner (1982) 
(see also eqs. 15, 16 of Harding \& Daugherty 1991 
and eq. 31 of NTZ08b) 
$$
\Gamma_{+} \approx 2\alpha B^3/3,
\eqno(A15a)
$$
$$
\Gamma_{-} \approx 4\alpha B^2/3,
\eqno(A15b)
$$
and substituting the above expressions 
into eq. (A11b) we get
$$
\Gamma_{mix} \approx \dfrac{4}{3}\alpha B^2 \sqrt{B/2},
\eqno(A15c)
$$
where $\alpha$ is the fine structure constant. For a discussion
regarding the cyclotron line widths see our Appendix C.
  
Note that the classical cross sections (1) 
and the expressions (4) are proportional 
to $r_0 c $, while the quantum-mechanical 
ones are proportional to $\sigma_T$. 
This means that the Lorentz profiles 
with variable the photon frequency 
(i.e. $L_{+}(\omega,\omega_r)$, $L_{-}(\omega,\omega_r)$ 
\& $L_{mix}(\omega,\omega_r)$) have 
dimensions of time and their relation
with the dimensionless ones
(i.e. ${\cal L}_{+}(\epsilon, \epsilon_r)$, 
${\cal L}_{-}(\epsilon, \epsilon_r)$ \& 
${\cal L}_{mix}(\epsilon, \epsilon_r)$) 
are (see Appendix C) 
$$
{\cal L}_{+}(\epsilon, \epsilon_r) \approx
{ {m_ec^2} \over \hbar} L_{+}(\omega, \omega_r),
\eqno(A16a)
$$
$$
{\cal L}_{-}(\epsilon, \epsilon_r) \approx
{ {m_ec^2} \over \hbar} L_{-}(\omega, \omega_r),
\eqno(A16b)
$$
$$
{\cal L}_{mix}(\epsilon, \epsilon_r) \approx
{ {m_ec^2} \over \hbar} L_{mix}(\omega, \omega_r),
\eqno(A16c)
$$
where the Lorentz profiles with 
argument the photon frequency 
are given by 
$$
L_{+}(\omega, \omega_r)= 
{ {\Gamma^\prime_{+}/2\pi} \over {(\omega - \omega_r)^2 
+ (\Gamma^\prime_{+}/2)^2} },
\eqno(A17a)
$$
$$
L_{-}(\omega, \omega_r)= 
{ {\Gamma^\prime_{-}/2\pi} \over {(\omega - 
\omega_r)^2 + (\Gamma^\prime_{-}/2)^2} },
\eqno(A17b)
$$
$$
L_{mix}(\omega, \omega_r)= 
{ {\Gamma^\prime_{mix}/2\pi} \over {(\omega - \omega_r)^2 + 
(\Gamma^\prime_{mix}/2)^2} },
\eqno(A17c)
$$
and
$$
\Gamma^\prime_{+} = \dfrac{m_ec^2}{\hbar}\Gamma_{+},
\eqno(A18a)
$$
$$
\Gamma^\prime_{-} = \dfrac{m_ec^2}{\hbar}\Gamma_{-} 
\approx \Gamma= 4e^2\omega^2_c/(3m_ec^3),
\eqno(A18b)
$$
$$
\Gamma^\prime_{mix} = \dfrac{m_ec^2}{\hbar}\Gamma_{mix}.
\eqno(A18c)
$$

Using the above, one finds that eq. (A6) is 
approximated by the following one
\begin{multline*}
\dfrac{d\sigma_{ss^\prime}}{d\cos\theta^\prime} 
\approx \dfrac{3\pi\sigma_T}{2}\dfrac{m_ec^2}{\hbar}  W(B, \theta, \theta^\prime) \\ 
\times \left[{ {(T_+^{s \to s^\prime})^2}\over {\Gamma_{+} }} L_{+} + 
{ {(T_-^{s \to s^\prime})^2}\over {\Gamma_{-} }} L_{-} + 
2{{T_+^{s \to s^\prime}T_-^{s \to s^\prime}} \over {\Gamma_{mix}} } {{L_{+} L_{-}} \over {L_{mix}}} 
\right],
\\
\shoveleft{= 2\pi \dfrac{3\pi r_o c}{8}\left(\dfrac{16 \alpha}{3}\right)}  
W(B, \theta, \theta^\prime)
\\ 
\times \left[{ {(T_+^{s \to s^\prime})^2}\over {\Gamma_{+} }} L_{+} + 
{ {(T_-^{s \to s^\prime})^2}\over {\Gamma_{-} }} L_{-} + 
2{{T_+^{s \to s^\prime}T_-^{s \to s^\prime}} \over {\Gamma_{mix}} } 
{{L_{+} L_{-}} \over {L_{mix}}} \right],
\tag{A19}
\end{multline*}
where 
$$
W(B, \theta, \theta^\prime)=1-B\left(2\sin^2\theta + 
(\cos\theta - \cos\theta^\prime)^2 \right),
\eqno(A20)
$$
and we have used that 
$$
\epsilon \approx \epsilon_r \approx B.
\eqno(A21)
$$

The expressions for $T_+^{s \to s^\prime}$ and  
$T_-^{s \to s^\prime}$ 
have strong polarization dependence 
and, as we mentioned earlier,
are given in the Appendix of NTZ08b. 
In order to obtain equations (4), 
we will work separately for each pair $s, s^\prime$.

\subsection{Transition $1 \to 1$}

By expanding the terms 
$T_{+}^{1 \to 1}$, $T_{-}^{1 \to 1}$ in 
the small parameters  $\epsilon, \epsilon^\prime$, 
and $B$, and employing eqs. (A15) and (A21), 
one obtains the following approximations
$$
\dfrac{(T^{1 \to 1}_+)^2}{\Gamma_+} \approx 
\dfrac{3}{16\alpha} \left(\dfrac{B}{2}\right),
\eqno(A22a)
$$
\begin{multline*}
\dfrac{(T^{1 \to 1}_-)^2}{\Gamma_-} \approx 
\dfrac{3}{16\alpha} \cos^2\theta\cos^2\theta^\prime
\\
\times \bigg[1
- B\left(2 + \sin^2\theta + 
\dfrac{2\sin^2\theta^\prime\cos\theta}{\cos\theta^\prime} - 
\sin^2\theta^\prime \right)\bigg] ,
\tag{A22b}
\end{multline*}
$$
\dfrac{2T^{1 \to 1}_+T^{1 \to 1}_-}{\Gamma_{mix}}  \approx 
\dfrac{3}{16\alpha}
\sqrt{2B}\cos\theta\cos\theta^\prime.
\eqno(A22c)
$$

Equations (A22), along with (A19), 
lead us to derive the following approximation
for $d\sigma_{11}/d\cos\theta^\prime$
\begin{multline*}
{ {d \sigma_{11}} \over {d\cos\theta^\prime} } 
\approx 2\pi { {3\pi r_o c} \over {8} }\Big[
g^{1 \to 1} \cdot L_{-} + h^{1 \to 1} \cdot L_{+}\\
+ \sqrt{2B}\cos\theta\cos\theta^\prime 
\dfrac{L_{-} \cdot L_{+}}{L_{mix}}
\Big],
\tag{A23a}
\end{multline*}
where
\begin{multline*}
g^{1 \to 1}(\theta,\theta^\prime, B)=\cos^2\theta \cos^2\theta^\prime\bigg[1 - 
B\bigg(3\sin^2\theta - 
\sin^2\theta^\prime \\ + (\cos\theta - \cos\theta^\prime)^2 + 2  
+ 2\sin^2\theta^\prime\dfrac{\cos\theta}{\cos\theta^\prime} \bigg)\bigg],
\tag{A23b}
\end{multline*}
and
$$
h^{1 \to 1}(B)=\dfrac{B}{2}.
\eqno(A23c)
$$

\subsection{Transition $1 \to 2$}    
 
We apply the same methodology to all the other cases.
Hence, by expanding the terms $T_{+}^{1 \to 2}$, 
$T_{-}^{1 \to 2}$ in the small parameters 
$\epsilon, \epsilon^\prime$, and $B$, 
and using eqs. (A15) and (A21), 
one finds the following approximations
$$
\dfrac{(T^{1 \to 2}_+)^2}{\Gamma_+} \approx \dfrac{3}{16\alpha}  
\left(\dfrac{B}{2} \cos^2\theta^\prime\right),
\eqno(A24a)
$$
\begin{multline*}
\dfrac{(T^{1 \to 2}_-)^2}{\Gamma_-} \approx \dfrac{3}{16\alpha} \cos^2\theta \left[1
- B\left(2 + \sin^2\theta \right)\right] ,
\tag{A24b}
\end{multline*}
$$
\dfrac{2T^{1 \to 2}_+T^{1 \to 2}_-}{\Gamma_{mix}}  \approx 
\dfrac{3}{16\alpha}
\sqrt{2B}\cos\theta\cos\theta^\prime.
\eqno(A24c)
$$

Then, by substituting eqs. (A24) into eq. (A19) 
one gets the following approximation
\begin{multline*}
{ {d \sigma_{12}} \over {d\cos\theta^\prime} } 
\approx 2\pi { {3\pi r_o c} \over {8} }\Big[
g^{1 \to 2} \cdot L_{-} + h^{1 \to 2} \cdot L_{+} \\
+ \sqrt{2B}\cos\theta\cos\theta^\prime 
\dfrac{L_{-} \cdot L_{+}}{L_{mix}}
\Big],
\tag{A25a}
\end{multline*}
where
\begin{multline*}
g^{1 \to 2}(\theta,\theta^\prime,B)=\cos^2\theta\bigg[1 - 
B\bigg(3\sin^2\theta + 2 \\
+ (\cos\theta - \cos\theta^\prime)^2 \bigg)\bigg],
\tag{A25b}
\end{multline*}
and
$$
h^{1 \to 2}(\theta^\prime,B)=
\dfrac{B}{2}\cos^2\theta^\prime.
\eqno(A25c)
$$

\subsection{Transition $2 \to 1$}     

Similarly, by expanding the terms $T_{+}^{2 \to 1}$, 
$T_{-}^{2 \to 1}$ in the small parameters 
$\epsilon, \epsilon^\prime$, and $B$, and taking 
into account eqs. (A15) and (A21), 
one deduces the following approximations
$$
\dfrac{(T^{2 \to 1}_+)^2}{\Gamma_+} \approx 
\dfrac{3}{16\alpha} \left(\dfrac{B}{2}
\cos^2\theta\right),
\eqno(A26a)
$$
\begin{multline*}
\dfrac{(T^{2 \to 1}_-)^2}{\Gamma_-} \approx \dfrac{3}{16\alpha} \cos^2\theta^\prime
\\
\times \bigg[1
- B\left(2 - \sin^2\theta^\prime + 
2\sin^2\theta^\prime\dfrac{\cos\theta}{\cos\theta^\prime} \right)\bigg],
\tag{A26b}
\end{multline*}
$$
\dfrac{2T^{2 \to 1}_+T^{2 \to 1}_-}{\Gamma_{mix}}  \approx 
\dfrac{3}{16\alpha}
\sqrt{2B}\cos\theta\cos\theta^\prime.
\eqno(A26c)
$$

Using eqs. (A26) and (A19), one finds 
\begin{multline*}
{ {d \sigma_{21}} \over {d\cos\theta^\prime} } \approx 
2\pi { {3\pi r_o c} \over {8} }\Big[
g^{2 \to 1} \cdot L_{-} + h^{2 \to 1} \cdot L_{+} \\
+ \sqrt{2B}\cos\theta\cos\theta^\prime 
\dfrac{L_{-} \cdot L_{+}}{L_{mix}}
\Big],
\tag{A27a}
\end{multline*}
where
\begin{multline*}
g^{2 \to 1}(\theta,\theta^\prime,B)=\cos^2\theta^\prime\bigg[1 - 
B \bigg(2\sin^2\theta + (\cos\theta - 
\cos\theta^\prime)^2 \\+ 2 - \sin^2\theta^\prime + 
2\sin^2\theta^\prime\dfrac{\cos\theta}{\cos\theta^\prime} \bigg)\bigg],
\tag{A27b}
\end{multline*}
and
$$
h^{2 \to 1}(\theta,B)=\dfrac{B}{2}\cos^2\theta.
\eqno(A27c)
$$

\subsection{Transition $2 \to 2$}

Following the same procedure as above, 
for the terms $T_{+}^{2 \to 2}$, $T_{-}^{2 \to 2}$ 
and employing eqs. (A15) and (A21), one derives that
$$
\dfrac{(T^{2 \to 2}_+)^2}{\Gamma_+} \approx 
\dfrac{3}{16\alpha} \left(\dfrac{B}{2}
\cos^2\theta\cos^2\theta^\prime\right),
\eqno(A28a)
$$
$$
\dfrac{(T^{2 \to 2}_-)^2}{\Gamma_-} \approx \dfrac{3}{16\alpha}  \bigg(1 - 
2B\bigg),
\eqno(A28b)
$$
$$
\dfrac{2T^{2 \to 2}_+T^{2 \to 2}_-}{\Gamma_{mix}}  \approx 
\dfrac{3}{16\alpha}
\sqrt{2B}\cos\theta\cos\theta^\prime.
\eqno(A28c)
$$

Thus, the substitution of eqs. (A28) into (A19), 
yields 
\begin{multline*}
{ {d \sigma_{22}} \over {d\cos\theta^\prime} } \approx 
2\pi { {3\pi r_o c} \over {8} }\Big[
g^{2 \to 2} \cdot L_{-} + h^{2 \to 2} \cdot L_{+} \\
+ \sqrt{2B}\cos\theta\cos\theta^\prime \dfrac{L_{-} 
\cdot L_{+}}{L_{mix}}
\Big],
\tag{A29a}
\end{multline*}
where
$$
g^{2 \to 2}(\theta,\theta^\prime,B)=1 - 
B\left[2\sin^2\theta + 
(\cos\theta - \cos\theta^\prime)^2 + 2 \right],
\eqno(A29b)
$$
and
$$
h^{2 \to 2}(\theta,\theta^\prime,B)=
\dfrac{B}{2}\cos^2\theta\cos^2\theta^\prime.
\eqno(A29c)
$$

\section{Approximations resulting from the Sina (1996) expressions}

We start from eq. (3.25) of Sina (1996),
written in the electron's rest 
frame (ground state), considering a unit 
system where $\hbar=m_e=c=1$, (i.e. 
energy is measured in $m_ec^2$, frequency 
is measured in $m_ec^2/\hbar$ and 
the magnetic field strength ${\cal B}$ is 
measured in ${\cal B}_{cr}=e^{-1}$) 
and neglect terms that are related 
to the 2nd Feynman diagram, as well 
as the terms that the intermediate electron 
Landau state has a principal 
quantum number $n\neq1$ (i.e. keeping only 
the terms that exhibit a divergence 
near resonant frequency with $n=1$, 
as we did in Appendix A). So, the 
infinite 
sum over $n$ for all possible 
intermediate states collapses 
to the 
following 
$$
\dfrac{d\sigma_{ss^\prime}}{d \Omega_f}=\dfrac{\alpha^2}{1-
	\beta_f\cos\theta_f}\dfrac{\omega_f}{\omega_i}\left|Z_{c1}(n=1)\right|^2,
\eqno(B1a)
$$
where $Z_{c1}(n=1)$ is given by eq. (3.15) of 
Sina (1996) and is a sum over the possible 
electron spin orientations $s_n$ in the 
intermediate state (for spin-up $s_n=+1$ 
whereas for spin-down $s_n=-1$)
$$
Z_{c1}(n=1)=\sum_{s_n}
\dfrac{D^{f,n=1,s_f,s_n}(k_f)H^{n=1,i,s_n,s_i}(k_i)}{\omega_i + E_i - E_{n=1,1} + 
i\Gamma^{n=1}_{s_n}/2}.
\eqno(B1b)
$$

We retain the notation of Sina (1996), 
so $\omega_i$ is the incident photon 
frequency, $E_i$ is the electron's energy before 
scattering, $\theta_i$, $\theta_f$ are the 
incident and scattered photon angles with 
respect to the magnetic field direction, 
the indices $i$ and $f$ stand for the initial 
and final electron Landau states (in 
our case they are equal to 0, i.e. ground state), 
the $s_i$ and $s_f$ are the initial and final 
electron's spin orientations and are equal 
to -1 (since only spin-down is allowed in 
the ground state), and $k_i$, $k_f$ are 
the incident and scattered electron 
wavenumbers, respectively. Furthermore, 
$E_{n=1,1}$ ($E_1$, hereafter) is the 
electron's energy in the intermediate 
state with $n=1$ and is given by eq. (3.3) 
of Sina (1996) (see also eq. 3.1 of Schwarm 2017)
$$
E_1\equiv E_{n=1,1}=\sqrt{1 + \omega^2_i\cos^2\theta_i + 2B},
\eqno(B2)
$$
where a misprint has been corrected and 
we have substituted eq. (3.4) of Sina (1996) 
into eq. (3.3) of Sina (1996) and took into 
account that the initial electron momentum 
$p_i$ is equal to zero, since we work in 
the electron's rest frame. 
Moreover, $\beta_f$ can be written as $p_f/E_f$ 
(see eq. B18 of Gonthier et al. 2014),
where $E_f$ and $p_f$ are the final electron 
energy and parallel component of momentum 
(with respect to the magnetic field direction) 
and are calculated by imposing the conservation 
of energy and by eq. (3.29) of Sina (1996). Thus
$$
E_f=1 + \omega_i - \omega_f,
\eqno(B3a)
$$
and
$$
p_f=\omega_i\cos\theta_i - \omega_f\cos\theta_f.
\eqno(B3b)
$$ 
Note that $\omega_f$ is the scattered photon 
frequency and is given by eq. (3.28) of Sina (1996), 
although it is actually our eq. (A5) written in 
a different form, so we do not present it here 
and $\Gamma^{n=1}_{s_n=+1}$, $\Gamma^{n=1}_{s_n=-1}$ 
($\Gamma_+$, $\Gamma_-$, hereafter) are the 
relativistic decay widths for the electron 
in the intermediate state with spin-up and 
spin-down, respectively, and are given 
by Herold, Ruder, \& Wunner (1982). Fig. 3.4 
of Schwarm (2017) verifies that the decay widths of Sina (1996) 
are in full agreement with the ones of Herold, Ruder, \& Wunner (1982) 
that we have used in Appendix A. This is reasonable 
and absolutely predictable, since both works have 
employed electron wavefunctions of Sokolov \& Ternov (1968).

Substituting the above into eqs. (B1) 
and using that the initial electron energy $E_1$ 
is equal to 1 and $\alpha^2=3\sigma_T/8\pi$ 
in this system of units, we obtain
\begin{multline*}
\dfrac{d\sigma_{ss^\prime}}{d \Omega_f}=\dfrac{3\sigma_T}{8\pi}\dfrac{1 + 
\omega_i - \omega_f}{1 + \omega_i - \omega_f -(\omega_i\cos\theta_i - 
\omega_f\cos\theta_f)\cos\theta_f} \\ \times \dfrac{\omega_f}{\omega_i}
\bigg|\dfrac{D^{f=0,n=1,s_f=-1,s_n=+1}(k_f)H^{n=1,i=0,s_n=+1,s_i=-1}(k_i)}{1 + 
\omega_i - E_{1} + i\Gamma_{+}/2} \\ + 
\dfrac{D^{f=0,n=1,s_f=-1,s_n=-1}(k_f)H^{n=1,i=0,s_n=-1,s_i=-1}(k_i)}{1 + 
\omega_i - E_{1} + i\Gamma_{-}/2}\bigg|^2,
\tag{B4}
\end{multline*}
where the complex functions $D^{f,n,s_f,s_n}(k_f)$ and
$H^{n,i,s_n,s_i}(k_i)$ have strong polarization 
dependence, since they depend on the polarization 
modes of the incident and scattered photons, 
and are given in Appendix D of Sina (1996). 
Specifically, the $D^{f,n,s_f,s_n}(k_f)$ terms 
exclusively depend on the final photon 
polarization mode and are calculated by 
eqs. (D.61), (D.66) of Sina (1996)
\begin{multline*}
D_{\perp}^{f=0,n=1,s_f=-1,s_n}(k_f)=\\i\bigg[\big(C_{1,f=0}C_{4,n=1}
+ C_{3,f=0}C_{2,n=1}\big)\Lambda_{-1,1}(k_f) -\\
\big(C_{2,f=0}C_{3,n=1} 
+ C_{4,f=0}C_{1,n=1}\big)\Lambda_{0,0}(k_f)\bigg],
\tag{B5a}
\end{multline*}
\begin{multline*}
D_{\parallel}^{f=0,n=1,s_f=-1,s_n}(k_f)=\\\cos\theta_f\bigg[\big(C_{1,f=0}C_{4,n=1}
+ C_{3,f=0}C_{2,n=1}\big)\Lambda_{-1,1}(k_f) +\\
\big(C_{2,f=0}C_{3,n=1} 
+ C_{4,f=0}C_{1,n=1}\big)\Lambda_{0,0}(k_f)\bigg] -\\
\sin\theta_f\bigg[\big(C_{1,f=0}C_{3,n=1}
+ C_{3,f=0}C_{1,n=1}\big)\Lambda_{-1,0}(k_f) -\\
\big(C_{2,f=0}C_{4,n=1} 
+ C_{4,f=0}C_{2,n=1}\big)\Lambda_{0,1}(k_f)\bigg],
\tag{B5b}
\end{multline*}  
while the terms $H^{n,i,s_n,s_i}(k_i)$  
exclusively depend on the initial photon 
polarization mode and are given by eqs. (D.60), (D.65) 
of Sina (1996)
\begin{multline*}
H_{\perp}^{n=1,i=0,s_n,s_i=-1}(k_i)=\\i\bigg[\big(C_{1,n=1}C_{4,i=0}
+ C_{3,n=1}C_{2,i=0}\big)\Lambda_{0,0}(k_i) -\\
\big(C_{2,n=1}C_{3,i=0} 
+ C_{4,n=1}C_{1,i=0}\big)\Lambda_{-1,1}(k_i)\bigg],
\tag{B6a}
\end{multline*}
\begin{multline*}
H_{\parallel}^{n=1,i=0,s_n,s_i=-1}(k_i)=\\
\cos\theta_i\bigg[\big(C_{1,n=1}C_{4,i=0}
+ C_{3,n=1}C_{2,i=0}\big)\Lambda_{0,0}(k_i) +\\
\big(C_{2,n=1}C_{3,i=0} 
+ C_{4,n=1}C_{1,i=0}\big)\Lambda_{-1,1}(k_i)\bigg] -\\
\sin\theta_i\bigg[\big(C_{1,n=1}C_{3,i=0}
+ C_{3,n=1}C_{1,i=0}\big)\Lambda_{-1,0}(k_i) -\\
\big(C_{2,n=1}C_{4,i=0} 
+ C_{4,n=1}C_{2,i=0}\big)\Lambda_{0,1}(k_i)\bigg],
\tag{B6b}
\end{multline*}
where the lower index ``$\parallel$" stands 
for the ordinary polarization mode whereas 
the lower index ``$\perp$" stands for 
the extraordinary polarization mode. 
Moreover the quantities $C_k$ with k=1,2,3,4, 
are the wavefunction coefficients of  
Sokolov \& Ternov (1968) and are given by 
eqs. (B61)-(B65) of Sina (1996), whereas the 
$\Lambda_{i,j}$ functions are proportional to 
Laguerre polynomials and are given in Appendix D 
of Sina (1996). Note that, one can also obtain 
all of the above using Appendix B of Gonthier et al. (2014).

After a lengthy, but straightforward calculation, 
we can write eq. (B4) in the following form
\begin{multline*}
\dfrac{d\sigma_{ss^\prime}}{d \cos\theta_f}=
\dfrac{3\pi\sigma_T}{2}
\dfrac{E_1}{\sqrt{1 + 2B\sin^2\theta_i}}\dfrac{\omega_f}{\omega_i} {\cal A}\\
\times
\left[{ {{\cal T}_+^{s \to s^\prime}}\over {\Gamma_{+} }} {\cal L}_{+} + 
{ {{\cal T}_-^{s \to s^\prime}}\over {\Gamma_{-} }} {\cal L}_{-} + 
2{{{\cal T}_{mix}^{s \to s^\prime}} \over {\Gamma_{mix}} } 
{{{\cal L}_{+}{\cal L}_{-}} \over {{\cal L}_{mix}}} \right],
\tag{B7a}
\end{multline*}
where we have done a trivial integration over the 
scattered photon azimuthal angle $\phi_f$. Also, for 
simplicity, we have defined the following 
quantities 
$$
{\cal A}=
\dfrac{(1 + \omega_i - \omega_f)\exp(-(\omega^2_i\sin^2\theta_i + 
\omega^2_f\sin^2\theta_f)/2B)}{1 + \omega_i - \omega_f - 
(\omega_i\cos\theta_i - \omega_f\cos\theta_f)\cos\theta_f},
\eqno(B7b)
$$
\begin{multline*}
{\cal T}_+^{s \to s^\prime}=\exp((\omega^2_i\sin^2\theta_i + \omega^2_f\sin^2\theta_f)/2B)\\
\times\left|D_{s^\prime}^{f=0,n=1,s_f=-1,s_n=+1}H_{s}^{n=1,i=0,s_n=+1,s_i=-1}\right|^2,
\tag{B7c}
\end{multline*}
\begin{multline*}
{\cal T}_-^{s \to s^\prime}=\exp((\omega^2_i\sin^2\theta_i + \omega^2_f\sin^2\theta_f)/2B)\\
\times\left|D_{s^\prime}^{f=0,n=1,s_f=-1,s_n=-1}H_{s}^{n=1,i=0,s_n=-1,s_i=-1}\right|^2,
\tag{B7d}
\end{multline*}
\begin{multline*}
2{\cal T}_{mix}^{s \to s^\prime}=\exp((\omega^2_i\sin^2\theta_i + \omega^2_f\sin^2\theta_f)/2B)\\
\times\bigg[\bigg((D_{s^\prime}^{f=0,n=1,s_f=-1,s_n=+1}H_{s}^{n=1,i=0,s_n=+1,s_i=-1}\big)^{*}\\
\times D_{s^\prime}^{f=0,n=1,s_f=-1,s_n=-1}H_{s}^{n=1,i=0,s_n=-1,s_i=-1} \bigg) +\\
\bigg(D_{s^\prime}^{f=0,n=1,s_f=-1,s_n=+1}H_{s}^{n=1,i=0,s_n=+1,s_i=-1}\\
\times\big(D_{s^\prime}^{f=0,n=1,s_f=-1,s_n=-1}H_{s}^{n=1,i=0,s_n=-1,s_i=-1} \big)^{*}\bigg)
\bigg].
\tag{B7e}
\end{multline*}

Note that the dimensionless Lorentz profiles 
${\cal L}_{+}(\omega_i, \omega_r)$, ${\cal L}_{-}(\omega_i, \omega_r)$, 
and ${\cal L}_{mix}(\omega_i, \omega_r)$ that are shown in 
eq. (B7a) are the same as the ones in 
Appendix A, since in this unit system the 
dimensionless photon frequencies $\omega_i, \omega_f$ 
and the dimensionless resonant frequency, have the 
same values with the corresponding dimensionless 
energies and, as we mentioned earlier, the decay 
widths are identical to those shown in Appendix A (and so are the effective decay widths given by eqs. A7a, A11b). 
Furthermore, the dimensionless resonant 
frequency $\omega_r$ is given by eq. (5) of 
Harding \& Daugherty (1991) and is actually 
the same as the dimensionless resonant energy (eq. A12) 
in Appendix A
$$
\omega_r=\dfrac{2B}{1 + \sqrt{1 + 2B\sin^2\theta_i}}.
\eqno(B8)
$$

Having obtained eqs. (B7), we are able 
to proceed to the derivation of eqs. (5a)-(5d).
Expansion up to first order 
in the small parameters $\omega_i$, $\omega_f$, and
$B$ yields
$$
{ \omega_f \over \omega_i} \approx 1-\dfrac{\omega_i}{2}(\cos\theta_i - \cos\theta_f)^2,
\eqno(B9)
$$
$$
\dfrac{\omega_f E_1\cdot{\cal A}}{\omega_i\sqrt{1 + 
2B\sin^2\theta_i}} \approx 1 - B\left(\sin^2\theta_i + \cos^2\theta_f - 2\cos\theta_i\cos\theta_f\right)
\eqno(B10)
$$
$$
\omega_i \approx \omega_r \approx B.
\eqno(B11)
$$

Using the above, one finds that eq. (B7a) 
can be approximated by
\begin{multline*}
\dfrac{d\sigma_{ss^\prime}}{d \cos\theta_f} \approx
\dfrac{3\pi\sigma_T}{2}{\cal W}(\theta_i,\theta_f,B)\\
\times
\left[{ {{\cal T}_+^{s \to s^\prime}}\over {\Gamma_{+} }} {\cal L}_{+} + 
{ {{\cal T}_-^{s \to s^\prime}}\over {\Gamma_{-} }} {\cal L}_{-} + 
2{{{\cal T}_{mix}^{s \to s^\prime}} \over {\Gamma_{mix}} } {{{\cal L}_{+}{\cal L}_{-}} \over {{\cal L}_{mix}}} 
\right],
\tag{B12}
\end{multline*}
and recalling the dimensions of each quantity 
by taking into account the procedure as 
well as the results of Appendix A, we obtain 
the following approximation
\begin{multline*}
\dfrac{d\sigma_{ss^\prime}}{d\cos\theta_f} \approx 
2\pi \dfrac{3\pi r_o c}{8}\left(\dfrac{16 \alpha}{3}\right) {\cal W}(\theta_i,\theta_f,B) \\ 
\times \left[{ {{\cal T}_+^{s \to s^\prime}}\over {\Gamma_{+} }} L_{+} + 
{ {{\cal T}_-^{s \to s^\prime}}\over {\Gamma_{-} }} L_{-} + 
2{{{\cal T}_{mix}^{s \to s^\prime}} \over {\Gamma_{mix}}}  {{L_{+} L_{-}} \over {L_{mix}}} 
\right],
\tag{B13a}
\end{multline*}
where 
$$
{\cal W}(\theta_i, \theta_f, B)=1-B\left(\sin^2\theta_i + 
\cos^2\theta_f - 2\cos\theta_i\cos\theta_f \right).
\eqno(B13b)
$$
From now on, all the quantities have their 
physical dimensions and as a result the Lorentz 
profiles that are shown in eq. (B13a) have 
dimensions of time and are calculated by eqs. (A17), 
where, in the notation that we use in this Appendix, 
the arguments of $L_{+}(\omega_i,\omega_r)$, 
$L_{i}(\omega_i,\omega_r)$, and $L_{mix}(\omega_i,\omega_r)$ are 
$\omega_i$ and $\omega_r$. As we said earlier,
the terms ${\cal T}_{+}^{s \to s^\prime}$, 
${\cal T}_{-}^{s \to s^\prime}$, and ${\cal T}_{mix}^{s \to s^\prime}$ 
have strong polarization dependence and 
for this reason we are going to derive 
the approximations separately for each 
possible combination of the incident and 
scattered photon polarization modes.

\subsection{Transition $1 \to 1$}

By expanding the terms ${\cal T}_{+}^{1 \to 1}$, 
${\cal T}_{-}^{1 \to 1}$, and ${\cal T}_{mix}^{1 \to 1} $ 
in the small parameters  $\omega_i, \omega_f$, and $B$, 
and employing eqs. (A15) and (B11), one obtains the 
following approximations
$$
\dfrac{{\cal T}^{1 \to 1}_+}{\Gamma_+} \approx 
\dfrac{3}{16\alpha} \left(\dfrac{B}{2}\right),
\eqno(B14a)
$$
\begin{multline*}
\dfrac{{\cal T}_{-}^{1 \to 1}}{\Gamma_-} \approx 
\dfrac{3}{16\alpha} \cos^2\theta_i\cos^2\theta_f
\\
\times \bigg[1
- B\left(\sin^2\theta_i + 
\dfrac{2\sin^2\theta_f\cos\theta_i}{\cos\theta_f} - 
\sin^2\theta_f \right)\bigg] ,
\tag{B14b}
\end{multline*}
$$
\dfrac{2{\cal T}_{mix}^{1 \to 1}}{\Gamma_{mix}}  \approx 
\dfrac{3}{16\alpha}
\sqrt{2B}\cos\theta_i\cos\theta_f.
\eqno(B14c)
$$

Equations (B14) along with (B13) lead us 
to derive the following approximation
for $d\sigma_{11}/d\cos\theta_f$
\begin{multline*}
{ {d \sigma_{11}} \over {d\cos\theta_f} } \approx 
2\pi { {3\pi r_o c} \over {8} }\Big[
{\cal G}^{1 \to 1} \cdot L_{-} + h^{1 \to 1} \cdot L_{+}\\
+ \sqrt{2B}\cos\theta_i\cos\theta_f \dfrac{L_{-} \cdot L_{+}}{L_{mix}}
\Big],
\tag{B15a}
\end{multline*}
where
\begin{multline*}
{\cal G}^{1 \to 1}(\theta_i,\theta_f,B)=\cos^2\theta_i \cos^2\theta_f\bigg[1 - 
B\bigg(3\sin^2\theta_i - \sin^2\theta_f \\ + 
(\cos\theta_i - \cos\theta_f)^2 -1  + 
2\sin^2\theta_f\dfrac{\cos\theta_i}{\cos\theta_f} \bigg)\bigg],
\tag{B15b}
\end{multline*}
and
$$
h^{1 \to 1}(B)=\dfrac{B}{2}.
\eqno(B15c)
$$

\subsection{Transition $1 \to 2$} 

We apply the same methodology to all the other cases.
Hence, by expanding the terms ${\cal T}_{+}^{1 \to 2}$, 
${\cal T}_{-}^{1 \to 2}$, and ${\cal T}_{mix}^{1 \to 2} $ in 
the small parameters  $\omega_i, \omega_f$, and $B$, 
and employing eqs. (A15) and (B11), 
one finds the following approximations
$$
\dfrac{{\cal T}^{1 \to 2}_+}{\Gamma_+} \approx 
\dfrac{3}{16\alpha}  \left(\dfrac{B}{2} 
\cos^2\theta_f\right),
\eqno(B16a)
$$
$$
\dfrac{{\cal T}^{1 \to 2}_-}{\Gamma_-} \approx 
\dfrac{3}{16\alpha} \cos^2\theta_i \left(1
- B\sin^2\theta_i\right) ,
\eqno(B16b)
$$
$$
\dfrac{2{\cal T}_{mix}^{1 \to 2}}{\Gamma_{mix}}  \approx 
\dfrac{3}{16\alpha}
\sqrt{2B}\cos\theta_i\cos\theta_f.
\eqno(B16c)
$$

Then, by substituting eqs. (B16) into eq. (B13a) 
one gets the following approximation
\begin{multline*}
{ {d \sigma_{12}} \over {d\cos\theta_f} } \approx 
2\pi { {3\pi r_o c} \over {8} }\Big[
{\cal G}^{1 \to 2} \cdot L_{-} + h^{1 \to 2} \cdot L_{+} \\
+ \sqrt{2B}\cos\theta_i\cos\theta_f \dfrac{L_{-} \cdot L_{+}}{L_{mix}}
\Big],
\tag{B17a}
\end{multline*}
where
\begin{multline*}
{\cal G}^{1 \to 2}(\theta_i,\theta_f,B)=\cos^2\theta_i\bigg[1 - 
B\bigg(3\sin^2\theta_i -1 \\
+ (\cos\theta_i - \cos\theta_f)^2 \bigg)\bigg],
\tag{B17b}
\end{multline*}
and
$$
h^{1 \to 2}(\theta_f,B)=\dfrac{B}{2}\cos^2\theta_f.
\eqno(B17c)
$$

\subsection{Transition $2 \to 1$}     

Similarly, by expanding the terms ${\cal T}_{+}^{2 \to 1}$, 
${\cal T}_{-}^{2 \to 1}$, and ${\cal T}_{mix}^{2 \to 1} $ in 
the small parameters  $\omega_i, \omega_f$, and $B$, 
and taking into account eqs. (A15) and (B11), 
one deduces the following approximations
$$
\dfrac{{\cal T}_{+}^{2 \to 1}}{\Gamma_+} \approx 
\dfrac{3}{16\alpha} \left(\dfrac{B}{2}
\cos^2\theta_i\right),
\eqno(B18a)
$$
\begin{multline*}
\dfrac{{\cal T}_{-}^{2 \to 1}}{\Gamma_-} \approx 
\dfrac{3}{16\alpha} \cos^2\theta_f
\\
\times \bigg[1
- B\left( 2\sin^2\theta_f\dfrac{\cos\theta_i}{\cos\theta_f}- 
\sin^2\theta_f \right)\bigg],
\tag{B18b}
\end{multline*}
$$
\dfrac{2{\cal T}_{mix}^{2 \to 1}}{\Gamma_{mix}}  \approx 
\dfrac{3}{16\alpha}
\sqrt{2B}\cos\theta_i\cos\theta_f.
\eqno(B18c)
$$

Using eqs. (B18) and (B13a), one finds
\begin{multline*}
{ {d \sigma_{21}} \over {d\cos\theta_f} } \approx 2\pi { {3\pi r_o c} \over {8} }\Big[
{\cal G}^{2 \to 1} \cdot L_{-} + h^{2 \to 1} \cdot L_{+} \\
+ \sqrt{2B}\cos\theta_i\cos\theta_f \dfrac{L_{-} \cdot L_{+}}{L_{mix}}
\Big],
\tag{B19a}
\end{multline*}
where
\begin{multline*}
{\cal G}^{2 \to 1}(\theta_i,\theta_f,B)=\cos^2\theta_f\bigg[1 - 
B \bigg(2\sin^2\theta_i + (\cos\theta_i - 
\cos\theta_f)^2 \\-1 - \sin^2\theta_f + 
2\sin^2\theta_f\dfrac{\cos\theta_i}{\cos\theta_f} \bigg)\bigg],
\tag{B19b}
\end{multline*}
and
$$
h^{2 \to 1}(\theta_i,B)=\dfrac{B}{2}\cos^2\theta_i.
\eqno(B19c)
$$

\subsection{Transition $2 \to 2$}

Following the same procedure as above 
for the terms ${\cal T}_{+}^{2 \to 2}$, 
${\cal T}_{-}^{2 \to 2}$, and ${\cal T}_{mix}^{2 \to 2} $ 
and employing eqs. (A15) and (B11), one derives that
$$
\dfrac{{\cal T}_{+}^{2 \to 2}}{\Gamma_+} \approx 
\dfrac{3}{16\alpha} \left(\dfrac{B}{2}\cos^2\theta_i\cos^2\theta_f\right),
\eqno(B20a)
$$
$$
\dfrac{{\cal T}_{-}^{2 \to 2}}{\Gamma_-} \approx \dfrac{3}{16\alpha}  ,
\eqno(B20b)
$$
$$
\dfrac{2{\cal T}_{mix}^{2 \to 2}}{\Gamma_{mix}}  \approx 
\dfrac{3}{16\alpha}
\sqrt{2B}\cos\theta_i\cos\theta_f.
\eqno(B20c)
$$

Thus, the substitution of eqs. (B20) into (B13a), 
yields 
\begin{multline*}
{ {d \sigma_{22}} \over {d\cos\theta_f} } \approx 
2\pi { {3\pi r_o c} \over {8} }\Big[
{\cal G}^{2 \to 2} \cdot L_{-} + h^{2 \to 2} \cdot L_{+} \\
+ \sqrt{2B}\cos\theta_i\cos\theta_f \dfrac{L_{-} \cdot L_{+}}{L_{mix}}
\Big],
\tag{B21a}
\end{multline*}
where
$$
{\cal G}^{2 \to 2}(\theta_i,\theta_f,B)=1 - 
B\left[2\sin^2\theta_i + 
(\cos\theta_i - \cos\theta_f)^2 - 1 \right],
\eqno(B21b)
$$
and
$$
h^{2 \to 2}(\theta_i,\theta_f,B)=
\dfrac{B}{2}\cos^2\theta_i\cos^2\theta_f.
\eqno(B21c)
$$  

\section{Cyclotron line widths}

The spin dependent relativistic cyclotron widths $\Gamma_{+},\,\Gamma_{-}$ for the transitions 
from the 1st excited intermediate electron state to the fundamental $n=0$ Landau state 
are given by eq. (17) of Herold et al. (1982) or/and by eq. (3) of Pavlov et al. (1991). 
So, we do not present the complete formulae here. 

Known in the literature are the non-relativistic cyclotron line widths, which consist of the dominant terms and are valid for small magnetic fields (${\cal B} \ll {\cal B}_{cr}$) (see Herold et al. 1982; NTZ08b). 
$$
\Gamma_{+} \approx 2\alpha B^3/3,
\eqno(C1a)
$$
$$
\Gamma_{-} \approx 4\alpha B^2/3.
\eqno(C1b)
$$
For users who would like to have more accuracy, we provide below next order corrections in $B$.  These corrections have been derived from numerical fits to the relativistic transition rates of Herold et al. (1982).
$$
\Gamma_{+} \approx \dfrac{2}{3}\alpha B^3(1 - 2.9B),
\eqno(C2a)
$$
$$
\Gamma_{-} \approx \dfrac{4}{3}\alpha B^2(1 - 2.7B).
\eqno(C2b)
$$

In Fig. 8 we display the relativistic decay rates of Herold (1982), along with the non-relativistic ones 
and the expressions C2, which have a first order correction in $B$.

\begin{figure}[h]
	\centering
	\includegraphics[angle=0,width=9.0cm]{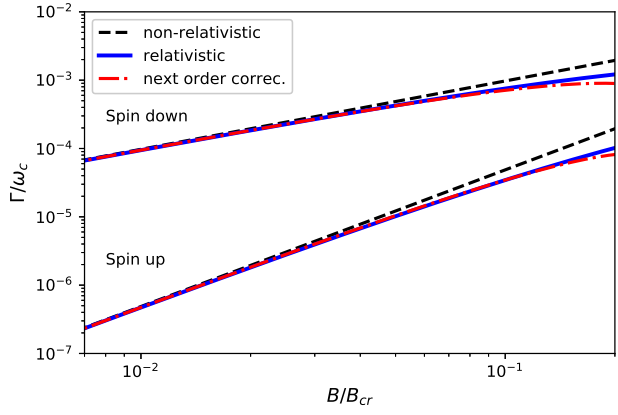}
	\caption{
		Comparison of the relativistic transition rates of Herold et al. (1982) (blue solid lines) 
		with the non-relativistic ones (eq. A15) (black dashed lines) and with
		the expressions given by eq. (C2) (red dash-dotted lines).	 Spin up refers to spin parallel to the magnetic field, whereas spin down refers to anti-parallel spin. The $\Gamma$'s are divided by the cyclotron frequency $\omega_c$ for presentation reasons. 
	}
	\label{Fig8}
\end{figure}

Clearly the non-relativistic expressions given by eq. (C1) can be used in the Lorentz profiles (defined by eq. A17) in the case where $B \ll 1$

\end{appendix}

\end{document}